\documentclass[12pt,thmsa]{article}
\usepackage{sw20lart}

\input tcilatex
\begin{document}

\author{Zet Gheorghe \and Department of Physics, Technical University ''Gh.Asachi''
\and Iasi, Romania}
\title{SELF-DUALITY EQUATIONS FOR SPHERICALLY SYMMETRIC SU(2) GAUGE FIELDS }
\date{May 14, 2001}
\maketitle

\begin{abstract}
A model of spherically symmetric SU(2) gauge theory is considered. The
self-duality equations are written and it is shown that they are compatible
with the Einstein-Yang-Mills equations.It is proven that this property is
true for any gauge theory with curved base space-time and having a compact
Lie group as structural group.
\end{abstract}

\section{Introduction}

The self-duality equations are important in gauge theories because they show
the connection between gauge models with internal symmetry groups and gauge
theory of gravity. They are differential equations of the first order and it
is easier to investigate the solutions for different particular
configurations of the gauge fields and of space-times.One of the most
important property of the self-duality equations is that they imply the
Yang-Mills field equations. In this paper we will prove this property for
the general case of a gauge theory with compact Lie group of symmetry over a
4-dimensional space-time manifold.

It is important to remark that there are $3m$ independent self-duality
equations (of the first order) while the number of Yang-Mills equations is
equal to $4m$, where $m$ is the dimension of the gauge group.Both of them
have $4m$ unknown functions which are the gauge potentials $A_\mu ^a(x),$ $%
a=1,2,....,m;$ $\mu =0,1,2,3$. But, we have, in addition, $m$ gauge
conditions for $A_\mu ^a(x)$ (for example Coulomb, Lorentz or axial gauge)
which together with the self-duality equation constitute a system of $4m$
equations. The Bianchi identities for the self-dual stress tensor $F_{\mu
\nu }^a$ coincide with the Yang-Mills equations and do not imply therefore
suplementary conditions.

In the sections 2 and 3 we will use the axial gauge in order to obtain the
self duality equations for a $SU(2)$ gauge theory over a cureved space-time.
The section 4 is devoted to the study of compatibility between self-duality
and Yang-Mills equations.In fact, we will write the Einstein-Yang-Mills
equations and we will analyse only the Yang-Mills sector. The Einstein
equations can not be obtained of course from self-duality. They should be
obtained if we would consider a gauge theory having $P\times SU(2)$ as
symmetry group, where $P$ is the Poincar\'{e} group. More generally, a gauge
theory of N-extended supersymmetry can be developed by imposing the
self-duality condition.

\section{The field-strength tensor of the gauge fields}

We will consider a $SU(2)$ gauge theory over a 4-dimensional space-time
manifold $M$, spherically symmetric, endowed with the metric: 
\begin{equation}
ds^2=\sigma ^2Ndt^2-\frac 1Ndr^2-r^2(d\theta ^2+\sin ^2\theta d\varphi ^2), 
\tag{2.1}
\end{equation}
where $\sigma $ and $N$ are function depending only of $r$ variable. For $%
\sigma =1$ and $N=1-\frac{2m}r$ we obtain the Schwarzschild metric, while
for $\sigma =1$ and $N=1-\frac{2M}r+\frac{Q^2+1}{r^2}$ we have the
Reissner-Nordstr\"{o}m (RS) metric. The components of the metric tensor $%
g_{\mu \nu }$ are: 
\begin{equation}
g_{00}=\sigma ^2N,\text{ }g_{11}=-\frac 1N,\text{ }g_{22}=-r^2,\text{ }%
g_{33}=-r^2\sin ^2\theta ,  \tag{2.2}
\end{equation}
and the determinant of this tensor is 
\begin{equation}
g=\det (g_{\mu \nu })=-\sigma ^2r^4\sin ^2\theta ,\text{ }\sqrt{-g}=\sigma
r^2\sin \theta .  \tag{2.3}
\end{equation}

The Lie algebra of $SU(2)$ group is characterized by the structure
equations: 
\begin{equation}
\lbrack T_a,T_b]=\varepsilon _{abc}T_c;\text{ }a,b,c=1,2,3,  \tag{2.4}
\end{equation}
where $\varepsilon _{abc}$ is the Levi-Civita symbol of rank 3 with $%
\varepsilon _{123}=1$. The basic elements of the model are $(M,g_{\mu \nu
},A)$, where $M$ is the space-time manifold with the metric coefficients
given in (2.2), and the Lie algebra-valued 1-form is $A=A_\mu dx^\mu =A_\mu
^aT_adx^\mu $. The Lie algebra-valued 2-form of the gauge fields $A_\mu ^a$
is defined by: 
\begin{equation}
F=\frac 12F_{\mu \nu }dx^\mu \wedge dx^\nu =\frac 12F_{\mu \nu }^aT_adx^\mu
\wedge dx^\nu ,  \tag{2.5}
\end{equation}
where $F_{\mu \nu }^a$ are the components of the stress tensor: 
\begin{equation}
F_{\mu \nu }^a=\partial _\mu A_\nu ^a-\partial _\nu A_\mu ^a+\varepsilon
_{abc}A_\mu ^bA_\nu ^c.  \tag{2.6}
\end{equation}

The spherically symmetric $SU(2)$ gauge fields will be parametrized as
(Witten ansatz)$^1$: 
\begin{equation}
A=uT_3dt+w(T_2d\theta -T_1\sin \theta d\varphi )+T_3\cos \theta d\varphi , 
\tag{2.7}
\end{equation}
where $u$ and $w$ are function depending only of variable $r$. Using the
ansatz (2.7), we obtain the following non-null components of the stress
tensor: 
\begin{eqnarray}
F_{02}^1 &=&-uw,\text{ }F_{13}^1=-w^{\prime }\sin \theta ,  \tag{2.8} \\
F_{03}^2 &=&-uw\sin \theta ,\text{ }F_{12}^2=w^{\prime },  \nonumber \\
F_{03}^3 &=&-u^{\prime },\text{ }F_{23}^3=(w^2-1)\sin \theta ,  \nonumber
\end{eqnarray}
with $u^{\prime }=\frac{du}{dr}$ and $w^{\prime }=\frac{dw}{dr}$.

Now, we introduce the dual 2-form $*F$ (the symbol $"*"$ denoting the Hodge
dual map) whose components are defined by: 
\begin{equation}
\ast F_{\mu \nu }^a=\frac 12\sqrt{-g}\varepsilon _{\mu \nu \rho \lambda
}F^{a\rho \lambda },  \tag{2.9}
\end{equation}
where $\varepsilon _{\mu \nu \rho \lambda }$ is the Levi-Civita symbol of
rank 4 with $\varepsilon _{0123}=1$. The non-null components of $*F$ are: 
\begin{eqnarray}
\ast F_{02}^1 &=&\sigma Nw^{\prime },\text{ }*F_{13}^1=-\frac{uw\sin \theta 
}{\sigma N},  \tag{2.10} \\
\ast F_{03}^2 &=&\sigma Nw^{\prime }\sin \theta ,\text{ }*F_{12}^2=\frac{uw}{%
\sigma N},  \nonumber \\
\ast F_{01}^3 &=&\frac{\sigma (w^2-1)}{r^2},\text{ }*F_{23}^3=\frac{%
r^2u^{\prime }\sin \theta }\sigma .  \nonumber
\end{eqnarray}
In the next section we will impose the self-duality condition $*F=iF$ for
the 2-form $F$ and we will obtain the self-duality equations for the $SU(2)$
gauge fields. We will show that these equations are compatible with the
Einstein-Yang-Mills $(EYM)$ equations of the model. In fact, we will prove
that this property is true for any self-dual gauge theory with compact gauge
Lie group.

\section{Self-duality equations}

A self-dual (or anti-self-dual) form $T$ over a differential manifold $M$
can be constructed only if $M$ is of even dimension and the following
equation is satisfied$^2$: 
\begin{equation}
\ast *T=\lambda T;\text{ }rankT=\frac 12\dim M.  \tag{3.1}
\end{equation}
But, the dual map (or the Hodge-duality) has the property: 
\begin{eqnarray}
\ast *T &=&(-1)^{k(n-k)}T\text{ }(forEuclidean\text{ }metric),  \tag{3.2} \\
\ast *T &=&-(-1)^{k(n-k)}T\text{ }(forMinkowski\text{ }metric),  \nonumber
\end{eqnarray}
where $k$ is the rank of $T$ and $n$ is the dimension of $M$. This means
that the quantity $\lambda $ in (3.1) is constrained to very special values: 
\[
\pm T=**T=*(\lambda T)=\lambda ^2T; 
\]
that is 
\begin{eqnarray}
\lambda &=&\pm 1,\text{ }if\text{ }**T=T\text{ },\text{ }(Euclidean\text{ }%
metric),  \tag{3.3} \\
\lambda &=&\pm i,\text{ }if\text{ }**T=-T,\text{ }(Minkowski\text{ }metric).
\nonumber
\end{eqnarray}

In our model, the rank of $F$ is $k=2$ and the dimension of the space-time $%
M $ is $n=4$. Then, the self-duality condition is$^3$:

\begin{equation}
\ast F=iF  \tag{3.4}
\end{equation}
Now, if we introduce the components (2.8) and (2.10) in (3.4), we obtain the
following self-duality equations: 
\begin{eqnarray}
\sigma Nw^{\prime } &=&-iuw,  \tag{3.5} \\
\frac{r^2}\sigma u^{\prime } &=&i(w^2-1).  \nonumber
\end{eqnarray}
These equations are of the first order and they determine the gauge
potential functions $u(r)$ and $w(r)$. The metric coefficients $\sigma $ and 
$N$ are determined by the EYM equations which are of second order and, as we
will prove in the next section, they determine also the gauge potential
functions $u(r)$ and $w(r)$.

The simplest solution of the self-duality equations (3.5) coupled with
Einstein equations (see Sect.4) is$^1$: 
\begin{equation}
u=0,\text{ }w=\pm 1,\text{ }\sigma =1,\text{ }N=1-\frac{2m}r,  \tag{3.6}
\end{equation}
which describes the Schwarzschild metric and a pure gauge Yang-Mills field.
The simplest solution with a non-trivial gauge field describes colored black
holes$^4$: 
\begin{equation}
u=u(\infty )+\frac Qr,\text{ }w=0,\text{ }N=1-\frac{2M}r+\frac{Q^2+1}{r^2},%
\text{ }\sigma =1.  \tag{3.7}
\end{equation}
It corresponds to the Reissner-Nordstr\"{o}m (RN) metric with the electric
charge $Q$ and the unit magnetic charge. Many others solutions
(particle-like, sphaleron type, with $\Lambda $-term, stringy type, axially
symmetric etc.) are given by Volkov and Gal'tsov$^1$. Local solutions of the
static, spherically symmetric, EYM equations with $SU(2)$ gauge group are
studied by Zotov$^5$ on the basis of dynamical system methods. In this case
it is proven the existence of solutions with oscillating metric as well as
the existence of local solutions for all known formal series expansions.
Exact solutions for self-dual SU(2) gauge theory with axial symmetry are
given in our paper$^6$.

\section{Einstein-Yang-Mills equations}

The integral action of our model is: 
\begin{equation}
S_{EYM}=\int \left\{ -\frac 1{16\pi G}R-\frac 1{4Ke^2}TrF_{\mu \nu }F^{\mu
\nu }\right\} \sqrt{-g}d^4x,  \tag{4.1}
\end{equation}
where $e$ is the gauge coupling constant, $R$ is the scalar curvature
associated to $g_{\mu \nu }$ and $Tr(T_aT_b)=K\delta _{ab}$. For $G=SU(2)$
we choose $T_a=\frac 12\tau _a$ ($\tau _a$ being the Pauli matrices) and
then $K=\frac 12$. The gravitational constant $G$ is the only dimensionful
quantity in the action (the units $\hbar =c=1$ are understood).

Taking $\delta S_{EYM}=0$ with respect to $A_\mu ^a$ and $g_{\mu \nu }$
fields, we obtain the following general form of the EYM equations$^3$: 
\begin{equation}
\frac 1{\sqrt{-g}}\partial _\mu (\sqrt{-g}F^{a\mu \nu })+f_{bc}^aA_\mu
^bF^{c\mu \nu }=0,(Yang-Mills\text{ }equations),  \tag{4.2}
\end{equation}
where $f_{bc}^a=-f_{cb}^a$ are the structure constants of the gauge group,
and respectively 
\begin{equation}
R_{\mu \nu }-\frac 12g_{\mu \nu }=8\pi GT_{\mu \nu },\text{ }(\func{Ei}nstein%
\text{ }equations),  \tag{4.3}
\end{equation}
with the gauge-invariant stress-energy tensor 
\begin{equation}
T_{\mu \nu }=\frac 1{Ke^2}Tr\left( -F_{\mu \rho }F_\nu \text{ }^\rho +\frac
14F_{\rho \lambda }F^{\rho \lambda }\right) ,  \tag{4.4}
\end{equation}
Then, introducing (2.2) and (2.8) in (4.2) and (4.3) we obtain the EYM
equations of our model: 
\begin{equation}
\left( \sigma Nw^{\prime }\right) ^{\prime }=\frac \sigma {r^2}w\left(
w^2-1\right) -\frac 1{\sigma N}u^2w,  \tag{4.5}
\end{equation}
\begin{equation}
\left( \frac{r^2}\sigma u^{\prime }\right) ^{\prime }=\frac 2{\sigma N}uw^2,
\tag{4.6}
\end{equation}
\begin{equation}
\frac{\sigma ^{\prime }}\sigma =\frac 2rw^{\prime 2}+\frac 2{N^2\sigma
^2r}w^2u^2,  \tag{4.7}
\end{equation}
\begin{equation}
\frac 12\left( N^{\prime }r+N-1\right) +\frac{r^2}{2\sigma ^2}u^{\prime
2}+\frac 1{N\sigma ^2}u^2w^2+Nw^{\prime 2}+\frac 1{2r^2}\left( w^2-1\right)
^2=0,  \tag{4.8}
\end{equation}
where we used $K=\frac 12$ and $\frac{4\pi G}{e^2}=1$ units. These equations
admit the particular solutions given in (3.6) and (3.7).

Let us compare the self-duality equations (3.5) with the first two EYM
equations (4.5) and (4.6). If we take the derivatives with respect to $r$ of
the equations (3.5), then we obtain: 
\begin{eqnarray}
\left( \sigma Nw^{\prime }\right) ^{\prime } &=&-i\left( u^{\prime
}w+uw^{\prime }\right) ,  \tag{4.9} \\
\left( \frac{r^2}\sigma u^{\prime }\right) ^{\prime } &=&2iww^{\prime }. 
\nonumber
\end{eqnarray}
Now, if we replace $iw^{\prime }$ and $iu^{\prime }$ deduced from (3.5) into
the left-hand sides of (4.9), then we obtain the EYM equations (4.5) and
(4.6). Of course, the other two EYM equations (4.7) and (4.8) can not be
obtained from the self-duality equations of the gauge fields. This may be
possible if we develope a gauge theory with the gauge group $P\times SU(2)$,
where $P$ is the Poincar\'{e} group$^3$.

In fact, the above result is valid for any compact Lie group. Indeed, if we
start with the self-duality equations (2.9), written in the equivalent form$%
^3$ 
\begin{equation}
\ast F^{a\mu \nu }=-\frac 1{2\sqrt{-g}}\varepsilon ^{\mu \nu \rho \lambda
}F_{\rho \lambda }^a=iF^{a\mu \nu },  \tag{4.10}
\end{equation}
then we have 
\begin{equation}
\frac i{\sqrt{-g}}\partial _\mu \left( \sqrt{-g}F^{a\mu \nu }\right) =-\frac
1{2\sqrt{-g}}\left( \varepsilon ^{\mu \nu \rho \lambda }\partial _\mu
F_{\rho \lambda }^a\right) .  \tag{4.11}
\end{equation}
But, it can be verified by a direct calculus that the following property is
satisfied if the tensor $F_{\rho \lambda }^a$ is self-dual: 
\begin{equation}
\frac 1{2\sqrt{-g}}\left( \varepsilon ^{\mu \nu \rho \lambda }\partial _\mu
F_{\rho \lambda }^a\right) =if_{bc}^aA_\mu ^bF^{c\mu \nu }.  \tag{4.12}
\end{equation}
Now, if we introduce (4.12) into (4.11) then we obtain the Yang-Mills
equations (4.2). Therefore, from the self-duality equations (of the first
order) one obtains the Yang-Mills equations (of the second order) for any
gauge theory with compact Lie group.

\section{Concluding remarks}

In this paper we have considered a spherically symmetric SU(2) gauge theory
in a curved spacetime. The self-duality equations for the gauge fields have
been obtained and the connection with the Yang-Mills equations has been
investigated. We proved that Yang-Mills equations result from duality and
that this property is true for any compact Lie group.

The Einstein equations for the metric coefficients can not be obtained from
self-duality. They should be derived if we would consider a gauge theory
with the direct product group $P\times SU(2)$, where $P$ is the Poincar\'{e}
group. Because the N-extended super- groups include both the Poincar\'{e}
and an internal group of symmetry like SU(N), it is important to study the
self-duality equations for the supersymmetry.$^7$

\section{Rferences}

1. M.S. Volkov, D.V. Gal'tsov, Phys.Reports 319, 1 (1999).

2. B. Felsager, Geometry, particles and fields, Odense Univ. Press, 1981.

3. G. Zet, Self-duality equations of gauge theories (to be published).

4. F.A. Bais, R.J. Russel, Phys. Rev. D11, 2692 (1975).

5. M.Yu. Zotov, Dynamical system analysis for the Einstein-Yang-Mills
equations, gr-qc/9906024.

6. G. Zet, V. Manta, Mod. Phys. Lett. A16, 685 (2001).

7. E. Sezgin, Self-duality and supersymmetry, hep-th/9212092.

\end{document}